\documentclass[A4, 10 pt, conference]{ieeeconf}  
\usepackage{amsmath}
\usepackage{amssymb} 
\IEEEoverridecommandlockouts                              

\usepackage[latin1]{inputenc} 
\usepackage{fancyhdr}
\usepackage[hang]{caption2}
\usepackage[english]{babel}
\usepackage{color}

\usepackage{epsfig}
\usepackage{graphicx}
\usepackage[normalsize,it,hang]{subfigure}
\usepackage{amsfonts,amsmath,bm,hhline}
\usepackage{xcolor}
\newtheorem{remark}{Remark}

\title{\LARGE \bf
Bullwhip effect attenuation in supply chain management \\ via control-theoretic tools and short-term forecasts: A preliminary study \\ with an application to perishable inventories}

\author{Koussaila Hamiche$^{1}$, Michel Fliess$^{2,4}$, C\'{e}dric Join$^{3,4}$ 
and Hassane Aboua\"{\i}ssa$^{1}$
\thanks{$^{1}$LGI2A (EA 3926), Universit\'{e} d'Artois, 64200 B\'{e}thune, France. \newline
Emails: koussailahamiche@gmail.com, hassane.abouaissa@univ-artois.fr}
\thanks{$^{2}$LIX (CNRS, UMR 7161), \'Ecole polytechnique, 91128 Palaiseau, France. \newline
Email: Michel.Fliess@polytechnique.edu}
\thanks{$^{3}$CRAN (CNRS, UMR 7039), Universit\'{e} de Lorraine, BP 239, 54506 Vand{\oe}uvre-l\`{e}s-Nancy, France. \newline
Email: cedric.join@univ-lorraine.fr}
\thanks{$^{4}$AL.I.E.N., 7 rue Maurice Barr\`{e}s, 54330 V\'{e}zelise, France. \newline
Emails: \{michel.fliess, cedric.join\}@alien-sas.com}
}

\begin{document}

\maketitle
\thispagestyle{empty}
\pagestyle{empty}
\begin{abstract}
Supply chain management and inventory control provide most exciting examples of control systems with delays. Here, Smith predictors, model-free control and new time series forecasting techniques are mixed in order to derive an efficient control synthesis. Perishable inventories are also taken into account. The most intriguing ``bullwhip effect'' is explained and attenuated, at least in some important situations. Numerous convincing computer simulations are presented and discussed.   

\end{abstract}


%
\IEEEpeerreviewmaketitle

\section{Introduction}
\emph{Supply chain management} \cite{boz,dolgui}, or \emph{SCM}, \textit{i.e.}, the management of the flow of goods and services, is a salient subfield of \emph{operations research}. Tools from control engineering have been tested since the very beginning (see, \textit{e.g.}, \cite{simon,vassian}). Unavoidable dead times, often called \emph{lead times} or \emph{throughput times}, are due to phenomena like transport and production. Delay differential equations play therefore an increasing r\^ole  (see, \textit{e.g.}, \cite{abbou,hsr1,cholo,dejon,hamiche,igna1,igna2,loiseau,riddals1,riddals2,schwartz1,schwartz2,schwartz3,sipahi,warburton}, and the references therein). Several previous publications use a model, that is more or less analogous to the following delay system, which is derived via a straightforward conservation principle,
\begin{equation}\label{1}
\dot{y}(t) = \frak{k} u(t - L) - d(t),  \quad t \geq L
\end{equation}
where
\begin{itemize}
\item the output variable $y(t) \geq 0$ is the material inventory,
\item the control variable $u(t) \geq 0$ is the factory supply,
\item $d(t) \geq 0$ is the customer demand,
\item $L \geq 0$ is the throughput time, which is known,
\item $\frak{k} \gneqq 0$ is a yield parameter, which might be poorly known.
\end{itemize}
The above infinite-dimensional system looks easy. The numerous corresponding control strategies in the literature do not seem however to be entirely satisfactory. Let us grasp the difficulty via the two following points:
\begin{itemize}
\item System \eqref{1} is \emph{marginally stable} \cite{franklin}.
\item Assume, for instance, that we want to keep the inventory constant, \textit{i.e.}, $\dot{y}(t) = 0$. Equation \eqref{1} yields $u(t) = \frac{d(t + L)}{\frak{k}}$.
The future value $d(t + L)$ of the customer demand has to be ``guessed.'' No theoretical technique will ever produce rigorously accurate predictions. For simplicity's sake let us presume here a constant bias $\frak{b} \neq 0$, \textit{i.e.}, $\dot{y}(t) = \frac{\frak{b}}{\frak{k}}$. Thus $y(t) = \frac{\frak{b}}{\frak{k}} (t - L) + y(L)$, $t \geq L$. Its absolute value $\mid{y(t)}\mid$ becomes very large, at least from a purely mathematical standpoint, when $t \to + \infty$. This is a convincing illustration of the famous \emph{bullwhip effect} to which a huge literature has been devoted (see, \textit{e.g.}, \cite{chen,dolgui,lee,sucky,ude,wang}, and the references therein).
\end{itemize}
Applying to System \eqref{1} the Smith predictor \cite{smith}, which is basic for controlling delay systems (see, \textit{e.g.}, \cite{franklin,lunze}), yields:\footnote{Smith predictors were already employed a few times for supply chain management. See, \textit{e.g.}, \cite{cholo,igna1,igna2,hsr1}.}
\begin{equation}\label{2}
\dot{\hat{y}}(t + L) = \frak{k} u(t ) - \hat{d}(t + L)
\end{equation}
where $\hat{y}(t + L)$ and $\hat{d}(t + L)$ are respectively the estimates of $y$ and $d$ at time $t + L$. In order to obtain $\hat{d}(t + L)$, we need what may be viewed as a short-term forecast. Here it will be achieved via time series, as repeatedly in chain supply chain management (see, \textit{e.g.}, \cite{axsater,aviv,thomo}). We apply here, like \cite{hamiche2}, a new setting which was proven to be useful in financial engineering \cite{perp,agadir}, and for motorway traffic \cite{troyes} and solar energy  \cite{sol} management. From a control-theoretic standpoint, this communication might be one of the very few publications that are combining Smith predictors, advanced forecasting techniques, and model-free control \cite{ijc13} 
since Equation \eqref{1}, which is only a poor modeling, is mainly used for the purpose of computer simulations. Our result indicates how to mitigate the bullwhip effect at least in some cases.\footnote{The literature does not seem to contain any clear-cut definition of this effect.} Several convincing computer simulations are displayed.

Controlling perishable inventories\footnote{Fresh food for instance.} is a not only a natural extension but also a key topic (see, \textit{e.g.}, \cite{dolgui,nahmias,sachs}, and the references therein). In order to provide straightforward numerical simulations, replace according to  \cite{igna2}  System \eqref{1} by
\begin{equation}\label{3}
\dot{y}(t) = - \sigma y(t) + \frak{k} u(t - L) - d(t),  \quad t \geq L
\end{equation}
where $\sigma \gneqq 0$ might be poorly known. The same tools as before are employed. 

Our paper is organized as follows. Our main tools are briefly reviewed in Section \ref{tool}. Numerous computer simulations are presented in Section \ref{exp}. Some concluding remarks may be found in Section \ref{conc}.
\section{About our main tools}\label{tool}
\subsection{Forecasting via time series}\label{basics}
Our presentation differs quite a lot from the existing approaches, where other time series techniques are advocated (see, \textit{e.g.}, \cite{axsater,ann,brockwell,melard}, and the references therein).
\subsubsection{Time series}
Take the time
interval $[0, 1] \subset \mathbb{R}$ and introduce as often in
\emph{nonstandard analysis} (see, \textit{e.g.}, \cite{robinson,diener,lobry,perp}) the infinitesimal sampling ${\mathfrak{T}} = \{ 0 = t_0 < t_1 < \dots < t_\nu = 1 \}$
where $t_{i+1} - t_{i}$, $0 \leq i < \nu$, is {\em infinitesimal},
{\it i.e.}, ``very small.'' A
\emph{time series} $X(t)$ is a function $X: {\mathfrak{T}}
\rightarrow \mathbb{R}$.

A time series ${\mathcal{X}}: {\mathfrak{T}} \rightarrow \mathbb{R}$
is said to be {\em quickly fluctuating}, or {\em oscillating}, if,
and only if, the integral $\int_A {\mathcal{X}} dm$ is
infinitesimal, \textit{i.e.}, very small, for any \emph{appreciable} interval, \textit{i.e.}, an interval which is neither very small nor very large.

According to a theorem due to Cartier and Perrin
\cite{cartier} the following additive decomposition holds for any time series $X$, which satisfies a weak integrability condition,
\begin{equation}\label{decomposition}
X(t) = E(X)(t) + X_{\tiny{\rm fluctuation}}(t)
\end{equation}
where
\begin{itemize}
\item the \emph{mean}, or \emph{trend}, $E(X)(t)$ is ``quite smooth,''
\item $X_{\tiny{\rm fluctuation}}(t)$ is quickly fluctuating.
\end{itemize}
The decomposition \eqref{decomposition} is unique up to an
infinitesimal.

\subsubsection{Forecasting}\label{forecast}
Let us start with the first degree polynomial time function $p_{1} (\tau)
= a_0 + a_1 \tau$, $\tau \geq 0$, $a_0, a_1 \in \mathbb{R}$. Rewrite
thanks to classic operational calculus with respect to the variable $\tau$ (see, \textit{e.g.},
\cite{yosida}) $p_1$ as $P_1 = \frac{a_0}{s} +
\frac{a_1}{s^2}$. Multiply both sides by $s^2$:
\begin{equation}\label{5}
s^2 P_1 = a_0 s + a_1
\end{equation}
Take the derivative of both sides with respect to $s$, which
corresponds in the time domain to the multiplication by $- t$:
\begin{equation}\label{6}
s^2 \frac{d P_1}{ds} + 2s P_1 = a_0
\end{equation}
The coefficients $a_0, a_1$ are obtained via the triangular system
of equations (\ref{5})-(\ref{6}). We get rid of the time
derivatives, \textit{i.e.}, of $s P_1$, $s^2 P_1$, and $s^2 \frac{d
P_1}{ds}$, by multiplying both sides of Equations
(\ref{1})-(\ref{2}) by $s^{ - n}$, $n \geq 2$. The corresponding
iterated time integrals are low pass filters which attenuate the
corrupting noises. A quite short time window is sufficient for
obtaining accurate values of $a_0$, $a_1$. Note that estimating $a_0$ yields the mean.

The extension to polynomial functions of higher degree is
straightforward. For derivative estimates up to some finite order
of a given smooth function $f: [0, + \infty) \to \mathbb{R}$, take a
suitable truncated Taylor expansion around a given time instant
$t_0$, and apply the previous computations. Resetting  and utilizing
sliding time windows permit to estimate derivatives of various
orders at any sampled time instant.
See \cite{easy,mboup,sira} for more details.

Set the following forecast $X_{\text{est}}(t + \Delta T)$, where $\Delta T > 0$ is not too ``large,''
\begin{equation}\label{delta}
X_{\text{forecast}}(t + \Delta T) = E(X)(t) + \left[\frac{d E(X)(t)}{dt}\right]_e \Delta T
\end{equation}
where $E(X)(t)$ and $\left[\frac{d E(X)(t)}{dt}\right]_e$ are estimated like $a_0$ and $a_1$ above. Let us stress that what we predict is the mean and not the quick fluctuations (see also \cite{perp,agadir,sol}).

\subsection{Model-free control}
The indisputable practical successes (see \cite{bldg,bara}, the numerous references therein, and \cite{abou,afsi,toulouse,cloud,dallas,hong,madadi,menhour,nd,ramp,marine,mocop,roman,sim,telsang,zhao,zhou}) of model-free control \cite{ijc13} explains why it has already been summarized recently many times: \cite{alinea,andrea,bara,iste,menhour}. It will therefore not be repeated here. This choice permits moreover to present computer experiments more thoroughly. 

\section{Inventory control simulations}\label{exp}

\subsection{Classic situation}
\subsubsection{Model-based proportional feedback controller} 
\begin{figure*}[!ht]
\centering%
\subfigure[\footnotesize Control]
{\epsfig{figure=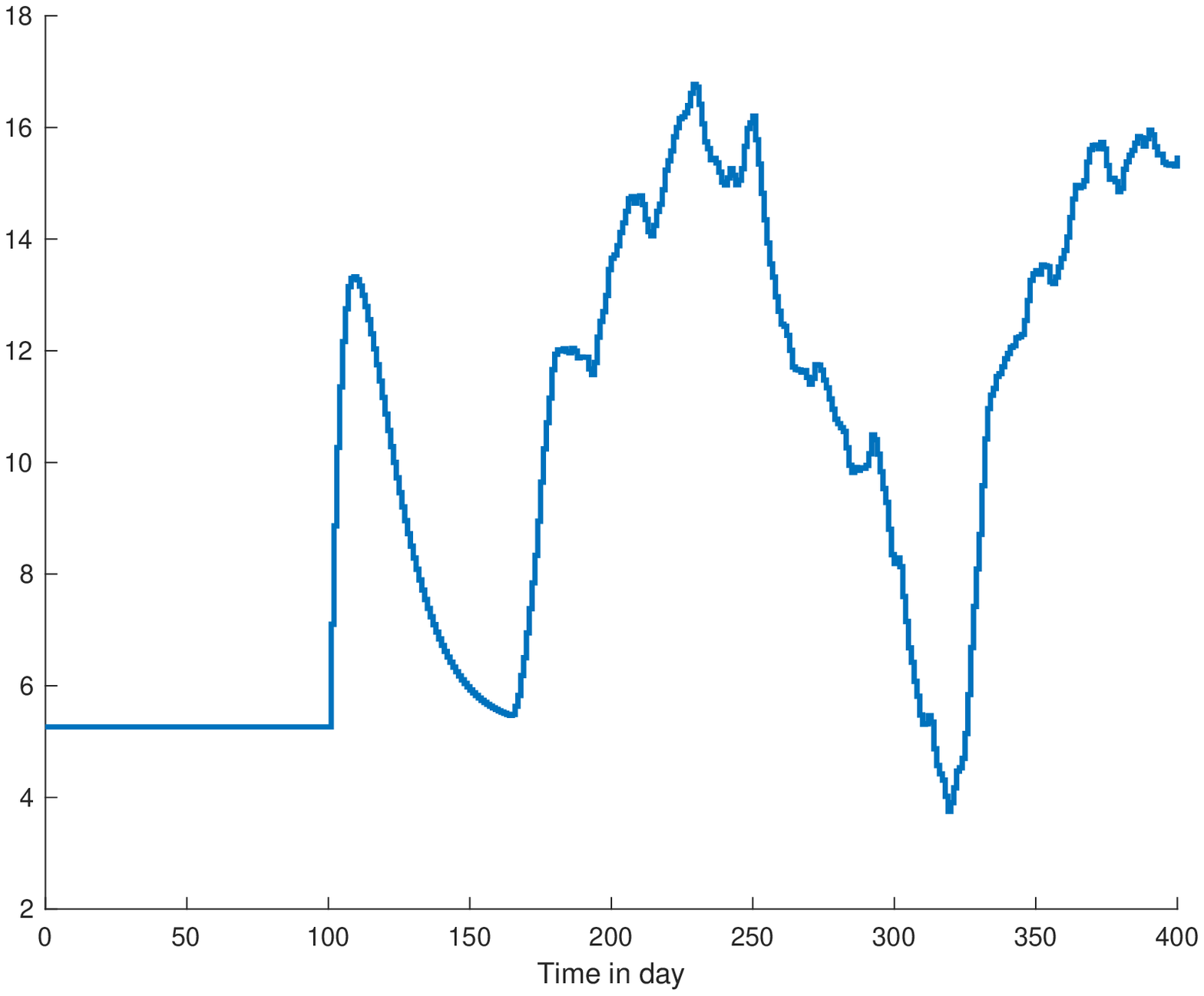,width=0.28\textwidth}}
\subfigure[\footnotesize Output (--) and Reference (- -)]
{\epsfig{figure=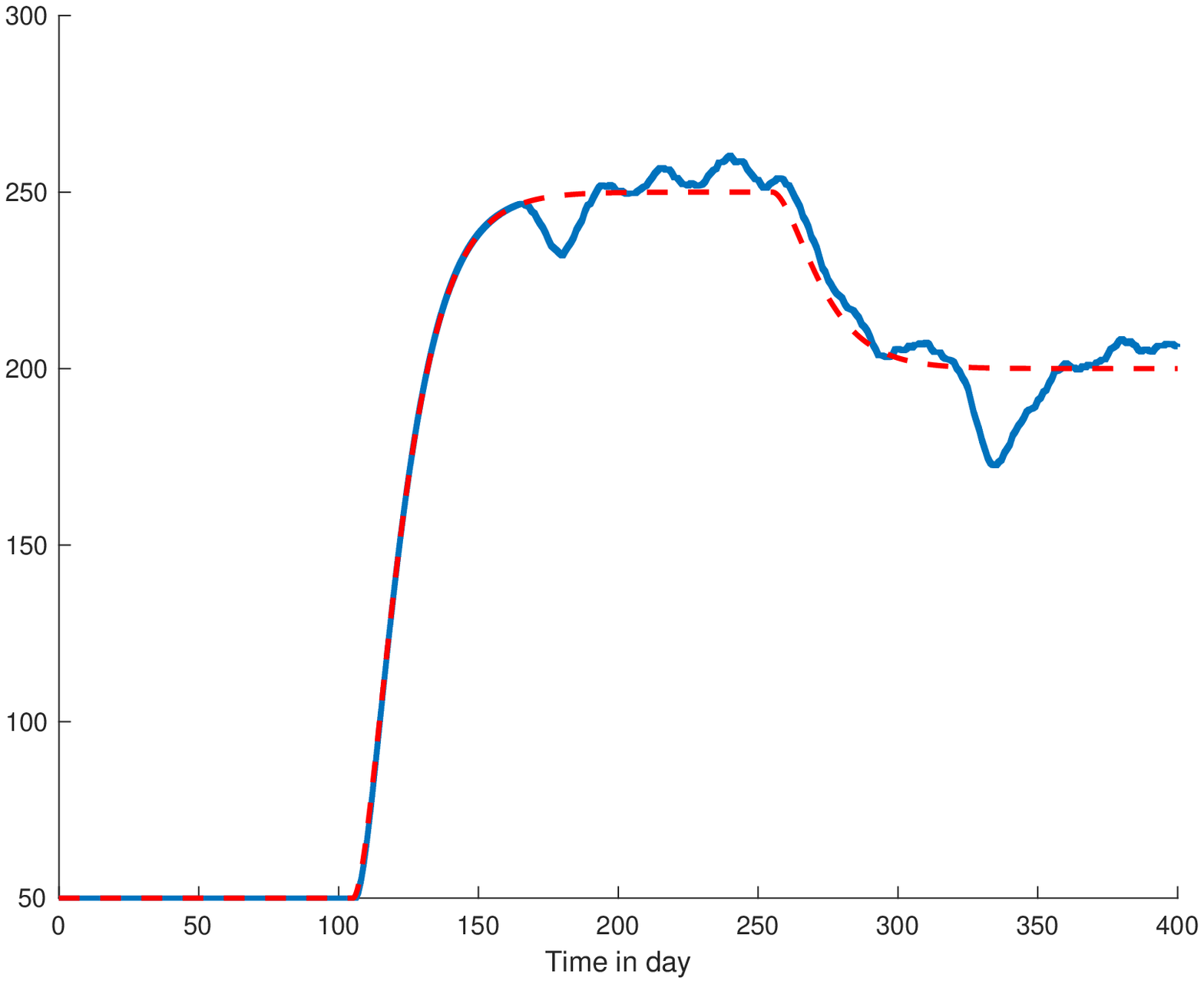,width=0.28\textwidth}}
\subfigure[\footnotesize Demand (--) and Demand forecasting (- -)]
{\epsfig{figure=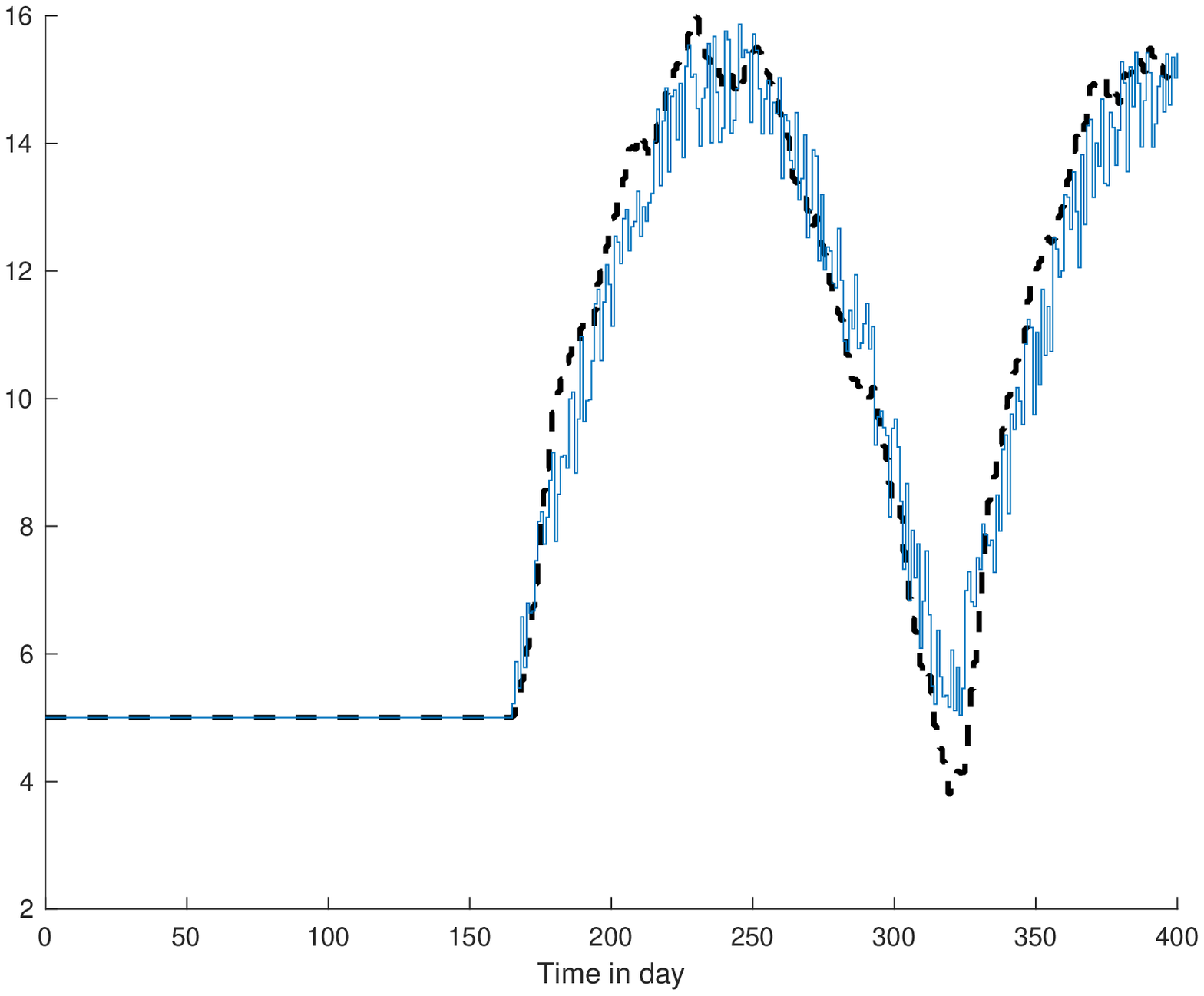,width=0.28\textwidth}}
\caption{Scenario 1}\label{S1}
\end{figure*}

The Smith predictor \eqref{2} stemming from Equation \eqref{1} yields the \emph{proportional controller} \cite{franklin}, or \emph{P controller},
\begin{equation}\label{cdeSmith}
u(t)=\frac{\dot y^\ast(t+L)+\hat d(t+L) - K_p \hat{e}(t+L)}{\frak{k}}
\end{equation}
where $y^\ast (t)$ is a reference inventory trajectory and $\hat{e}(t + L) = \hat y(t+L)-y^\ast(t+L)$ is the predicted tracking error. This last quantity is obtained thanks to the forecast $\hat{d}(t + L)$ of the customer demand via the techniques sketched in Section \ref{forecast}. Figure \ref{S1} displays excellent results with $K_p = 0.1$. The data are borrowed from \cite{schwartz1}. The customer demand is represented in Figure \ref{S1}-(c): note that after the constant portion at the beginning a corrupting uniform white noise $n$, $-1 \leq n \leq 1$,  has been added. The sampling time is $1$ day. 

\begin{figure*}[!ht]
\centering%
\subfigure[\footnotesize Control]
{\epsfig{figure=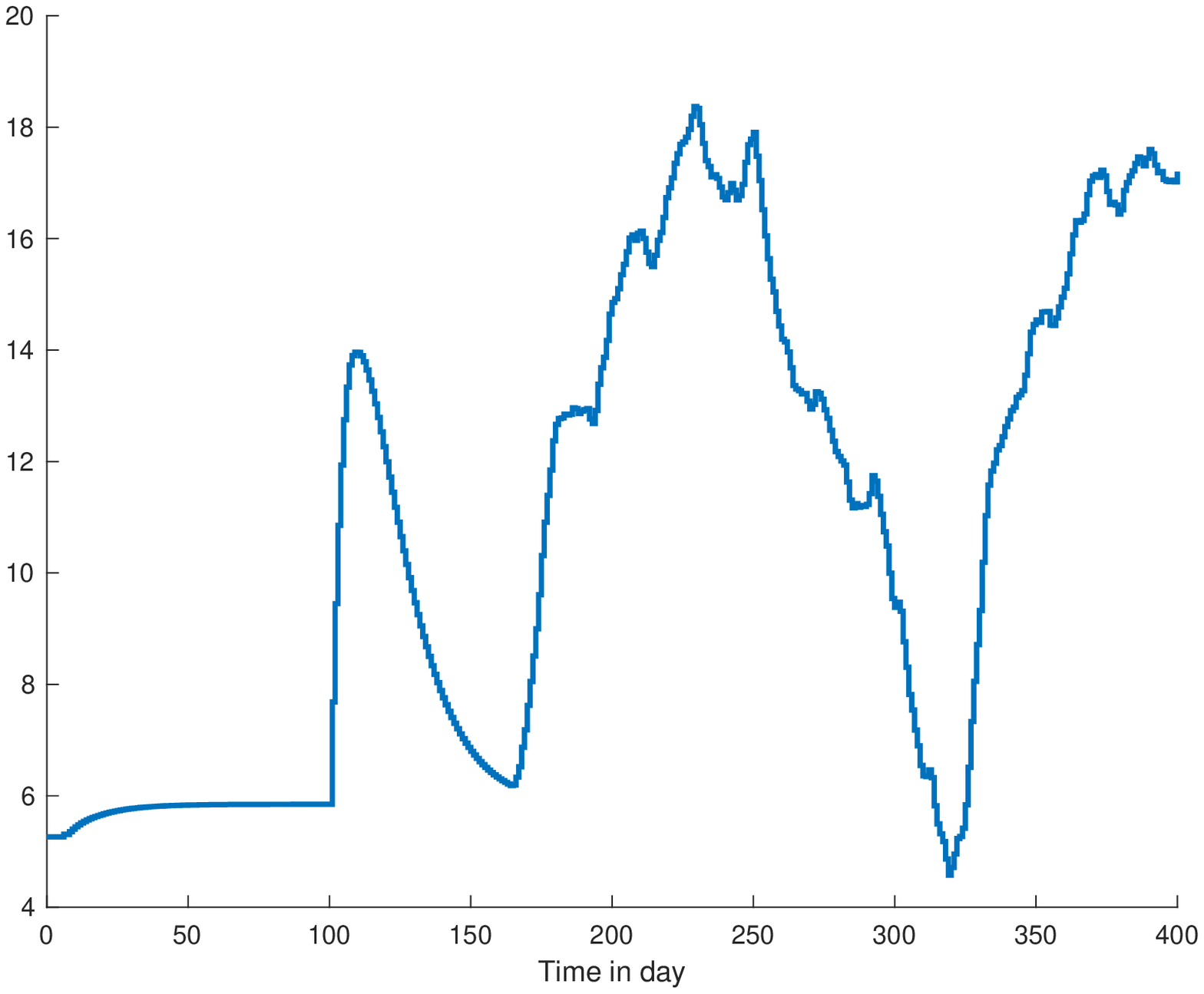,width=0.28\textwidth}}
\subfigure[\footnotesize Output (--) and Reference (- -)]
{\epsfig{figure=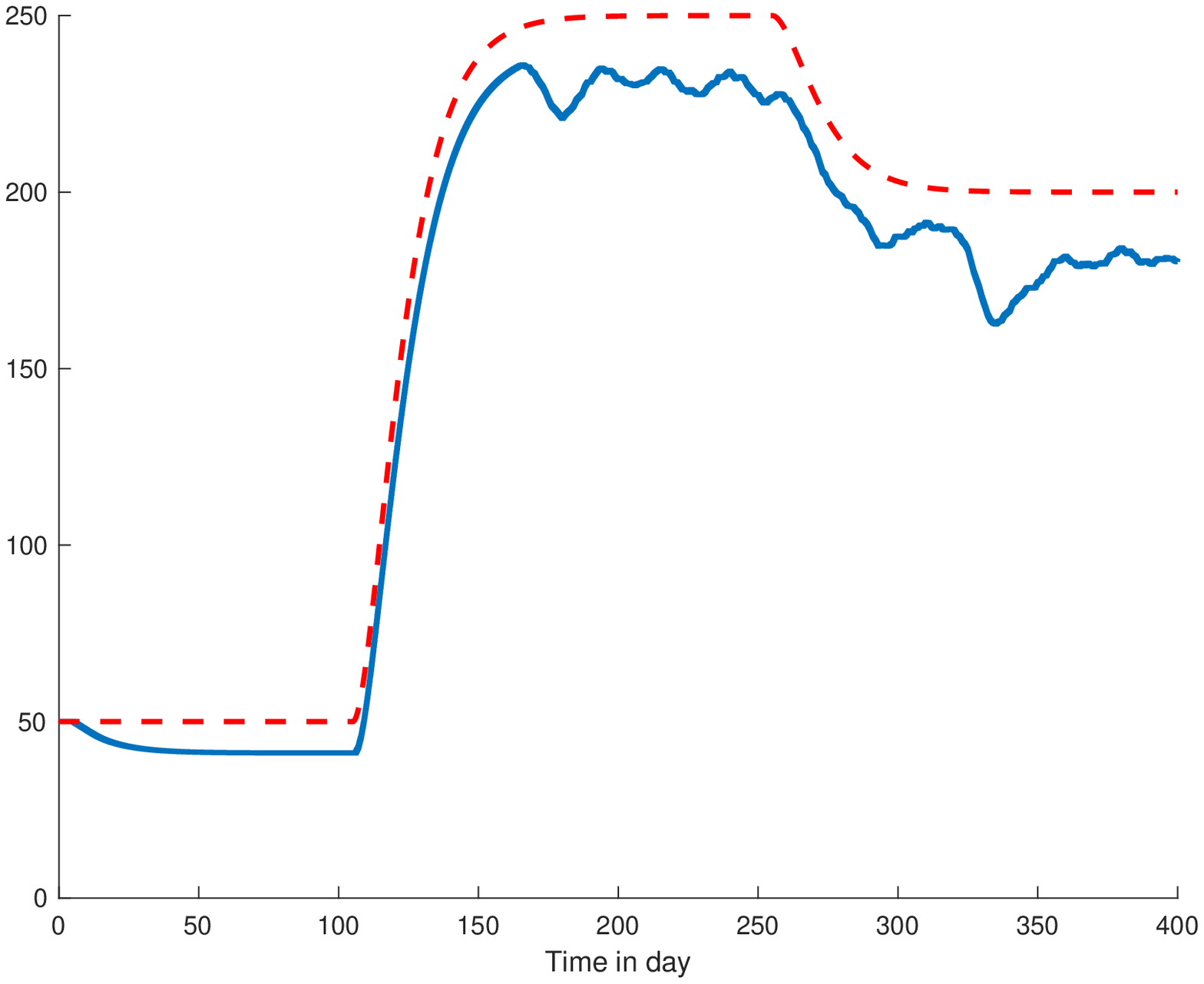,width=0.28\textwidth}}
\subfigure[\footnotesize Demand (--) and Demand forecasting (- -)]
{\epsfig{figure=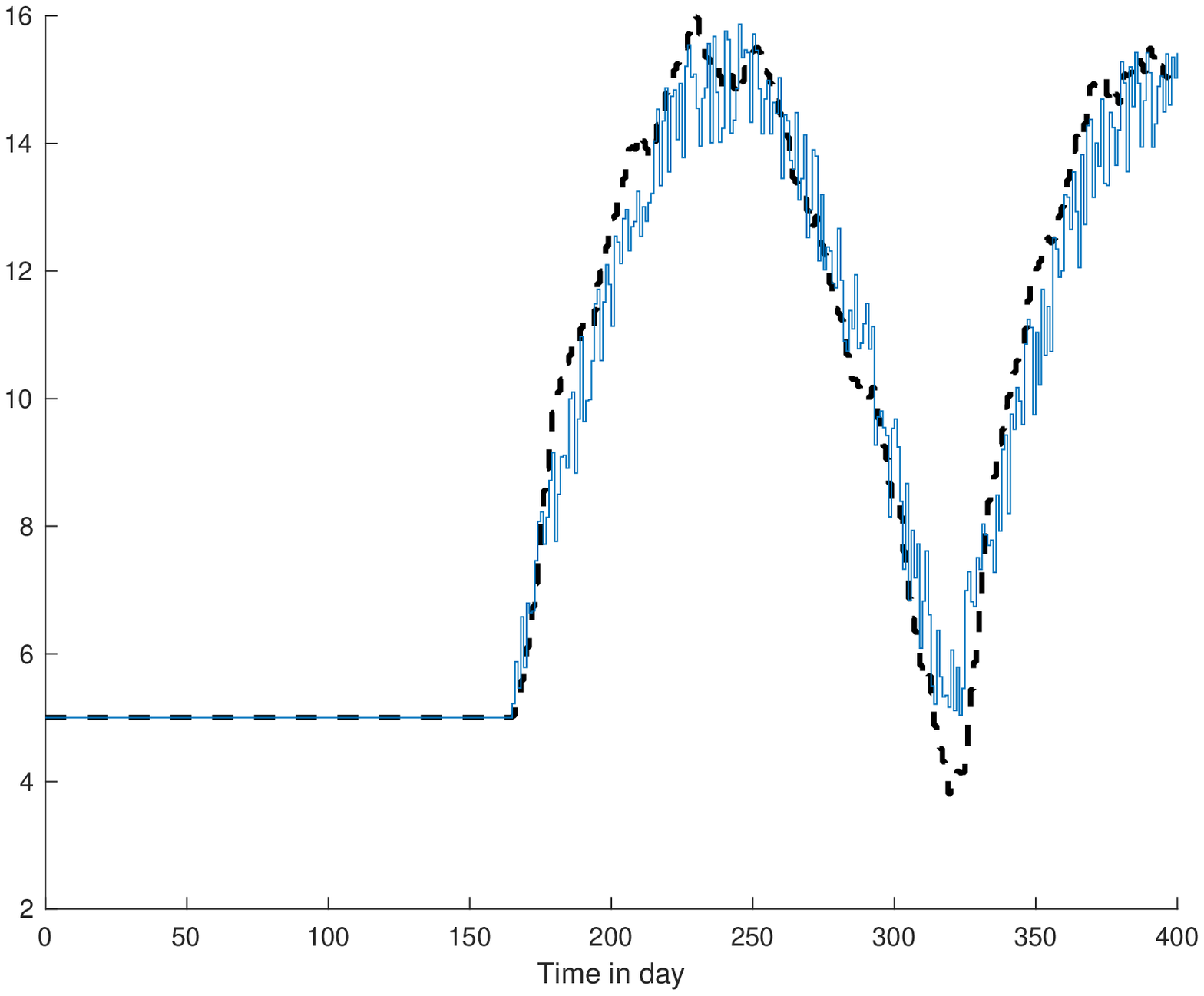,width=0.28\textwidth}}
\caption{Scenario 2}\label{S2}
\end{figure*}

In Equation \eqref{1}, $L = 5$, $\frak{k} = 0.95$.
With a poorly known model, \textit{i.e.}, $\frak{k}_{\text{model}}=0.95 \times 0.9 = 0.855$, a noticeable deterioration is shown in Figure \ref{S2} with the same assumptions, \textit{i.e.}, $\frak{k} = 0.95$.

\subsubsection{Model-free control and the corresponding intelligent controller}
\begin{figure*}[!htb]
\centering%
\subfigure[\footnotesize Control]
{\epsfig{figure=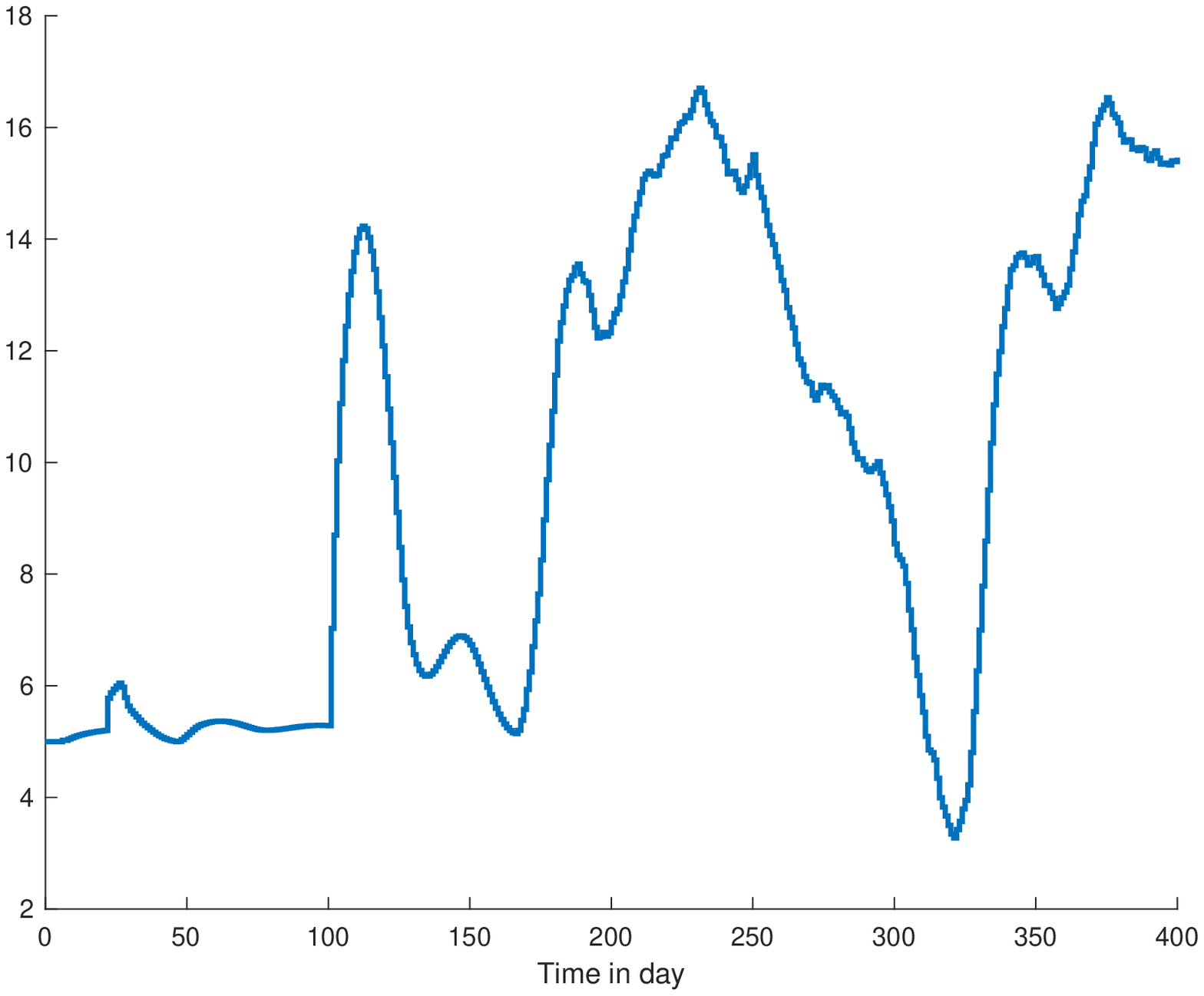,width=0.28\textwidth}}
\subfigure[\footnotesize Output (--) and Reference (- -)]
{\epsfig{figure=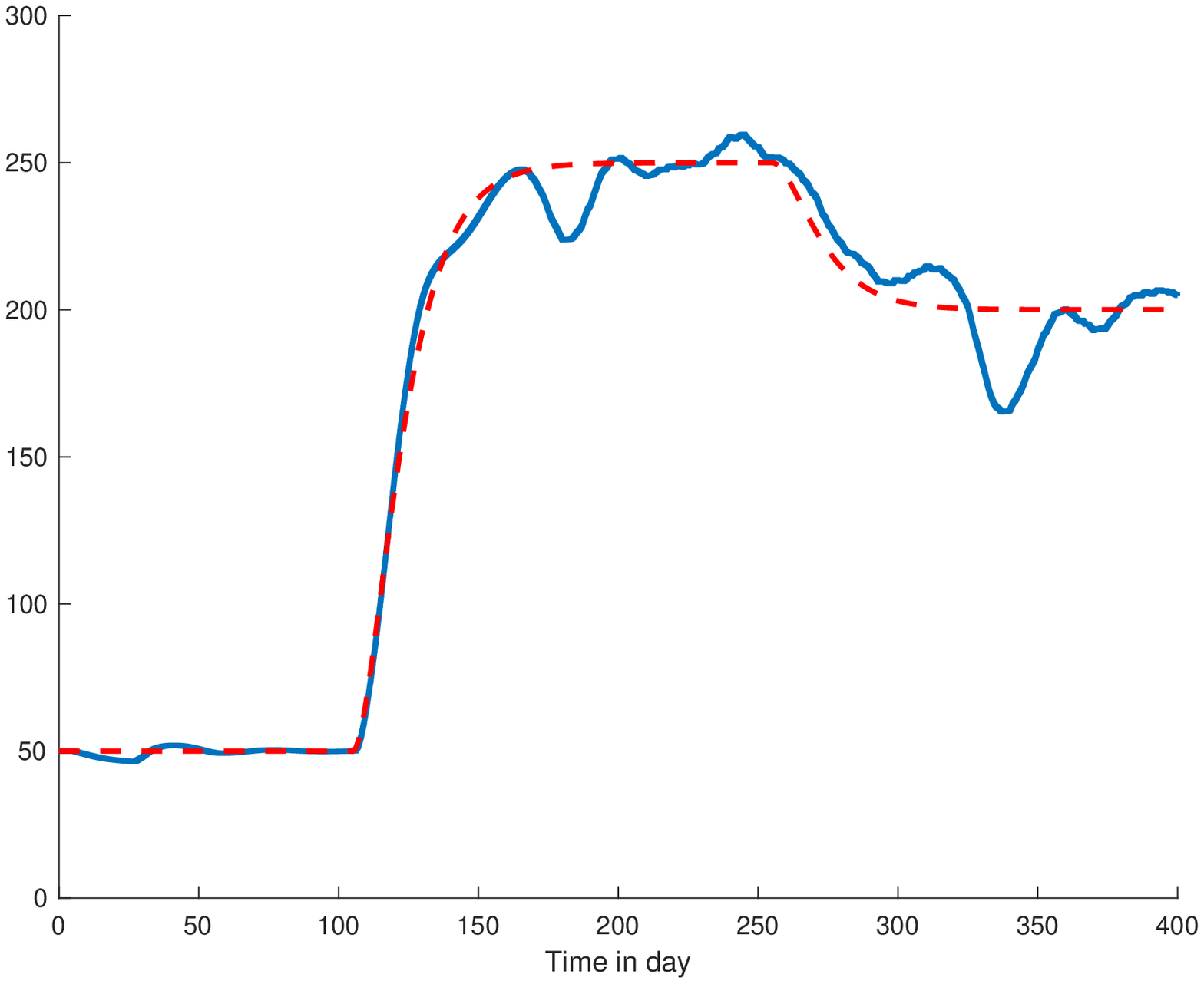,width=0.28\textwidth}}
%
%
\caption{Scenario 3}\label{S3}
\end{figure*}

The obvious analogue of the \emph{ultra-local model} in the model-free setting \cite{ijc13} reads here:
\begin{equation}\label{Ultra}
\dot y(t)=\alpha u(t-L) - F(t)
\end{equation}
where $F(t)$ subsumes the poorly known internal system structure as well as the external perturbations. The coefficient $\alpha$ is chosen by the practitioner in such a way that the three terms in Equation \eqref{Ultra} are of the same magnitude. It yields the corresponding \emph{intelligent proportional controller} \cite{ijc13}, or \emph{iP},
\begin{equation}\label{Ucsm}
u(t)=\frac{\dot y^\ast(t+L)+\hat{F}(t+L) - K_p \hat{e}(t+L)}{\alpha}
\end{equation}
The forecast $\hat{F}(t+L)$ of $F$ is again obtained via Section \ref{forecast}. With the same $\frak{k}_{\text{model}}=0.95 \times 0.9 = 0.855$ and the same $K_p = 0.1$, and with $\alpha = 1$, the results depicted in Figure \ref{S3} improve a lot when compared to Figure \ref{S2}.
\subsubsection{Bullwhip effect attenuation}\label{bull}
Figure \ref{S3}-(b) in particular shows the vanishing of the static error due to a poor knowledge of the model. According to the literature it should be viewed as an instance of bullwhip effect attenuation.

Assume now that the error due to the forecast is bounded by some quantity $M > 0$. Straightforward calculations show that the choice of a  \emph{high gain} (see, \textit{e.g.}, \cite{lunze}) $K_p \ggg 0$ minimizes the effect on $y$. Note
however that the oscillations of the tracking error $e$ are dangerously amplified on $u$ by such a choice, which should be therefore avoided as much as possible. An ``efficient'' forecasting technique is therefore of utmost importance.

\subsection{Perishable inventory}\label{PP}
Consider System \eqref{3}. Apply the iP \eqref{Ucsm}. Data are borrowed from \cite{igna2}. Choose $L = 7$, $\sigma=0.08$, and $\frak{k} = 1$. The sampling time is now $0.1$ day. 

\begin{figure*}[!ht]
\centering%
\subfigure[\footnotesize Control]
{\epsfig{figure=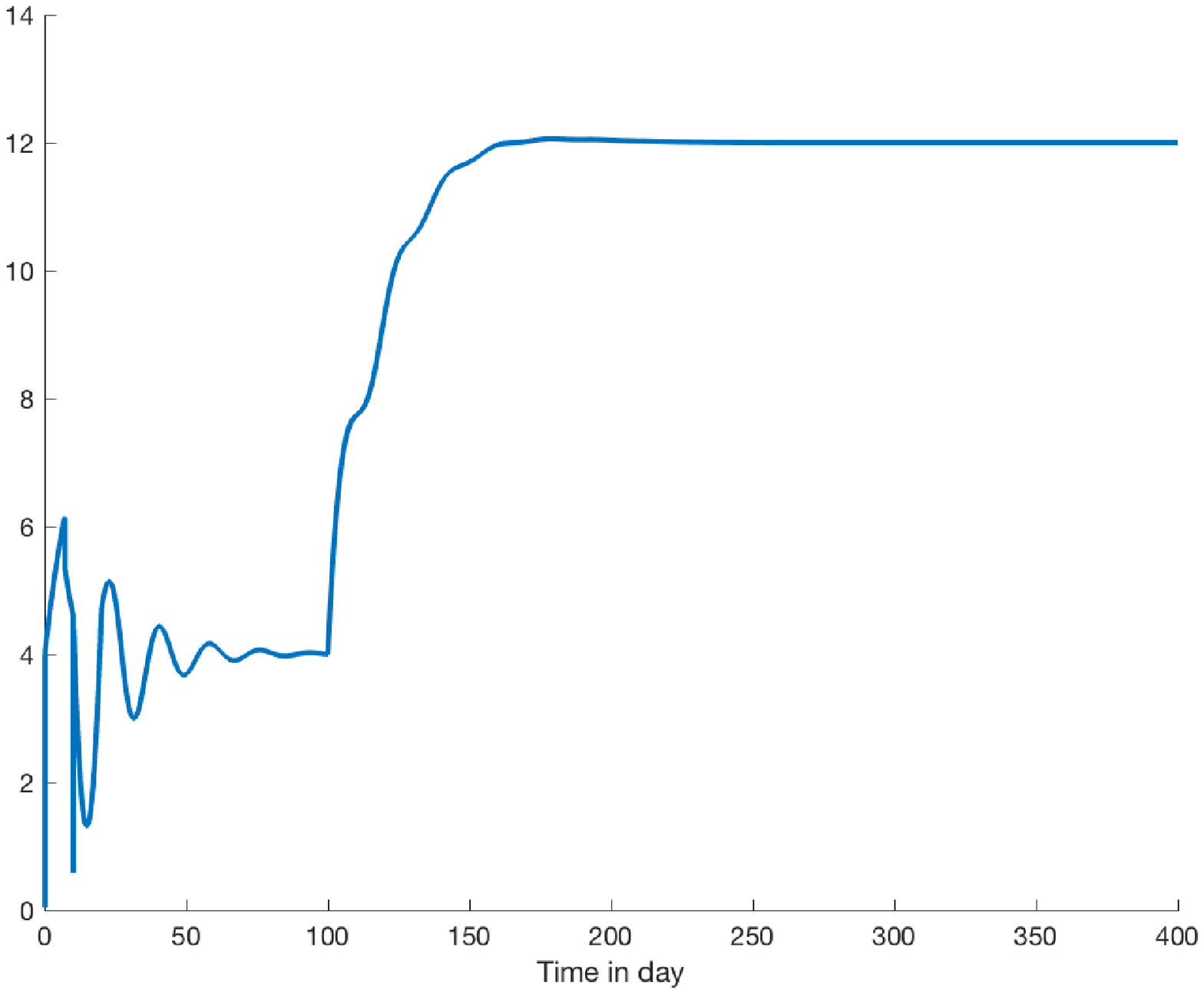,width=0.28\textwidth}}
\subfigure[\footnotesize Output (--) and Reference (- -)]
{\epsfig{figure=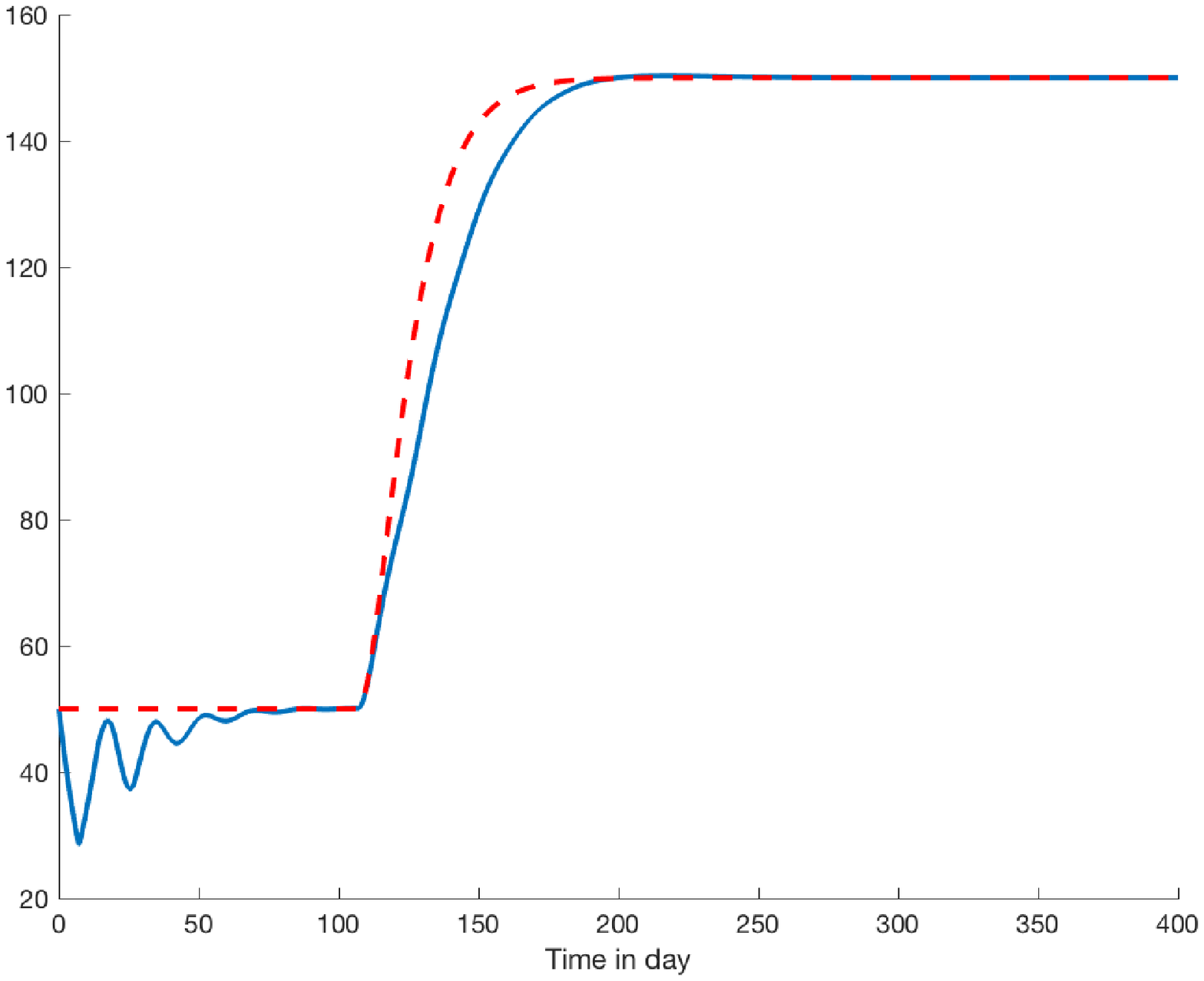,width=0.28\textwidth}}
%
%
\caption{Scenario 4}\label{S4}
\end{figure*}

\begin{figure*}[!ht]
\centering%
\subfigure[\footnotesize Control]
{\epsfig{figure=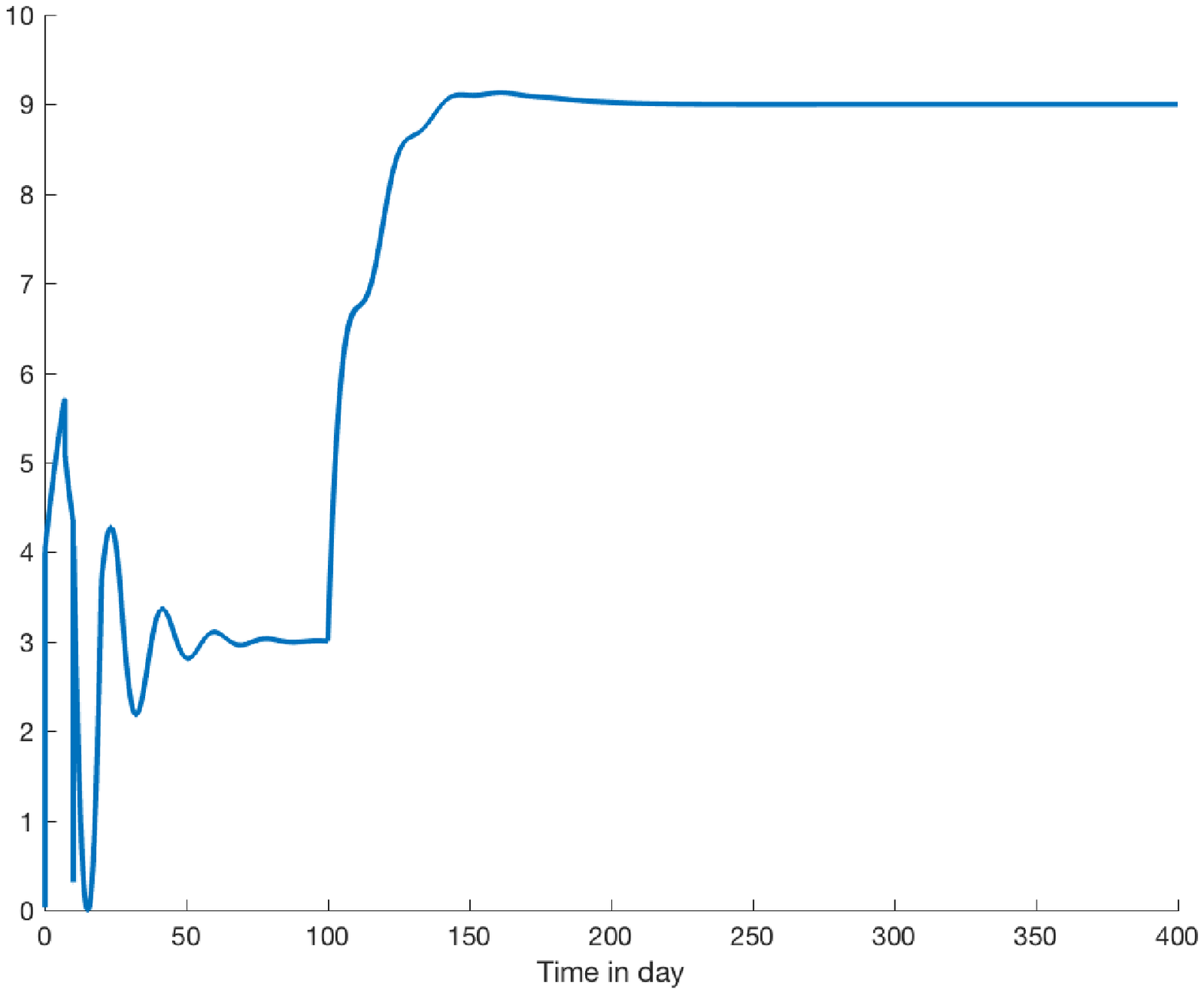,width=0.28\textwidth}}
\subfigure[\footnotesize Output (--) and Reference (- -)]
{\epsfig{figure=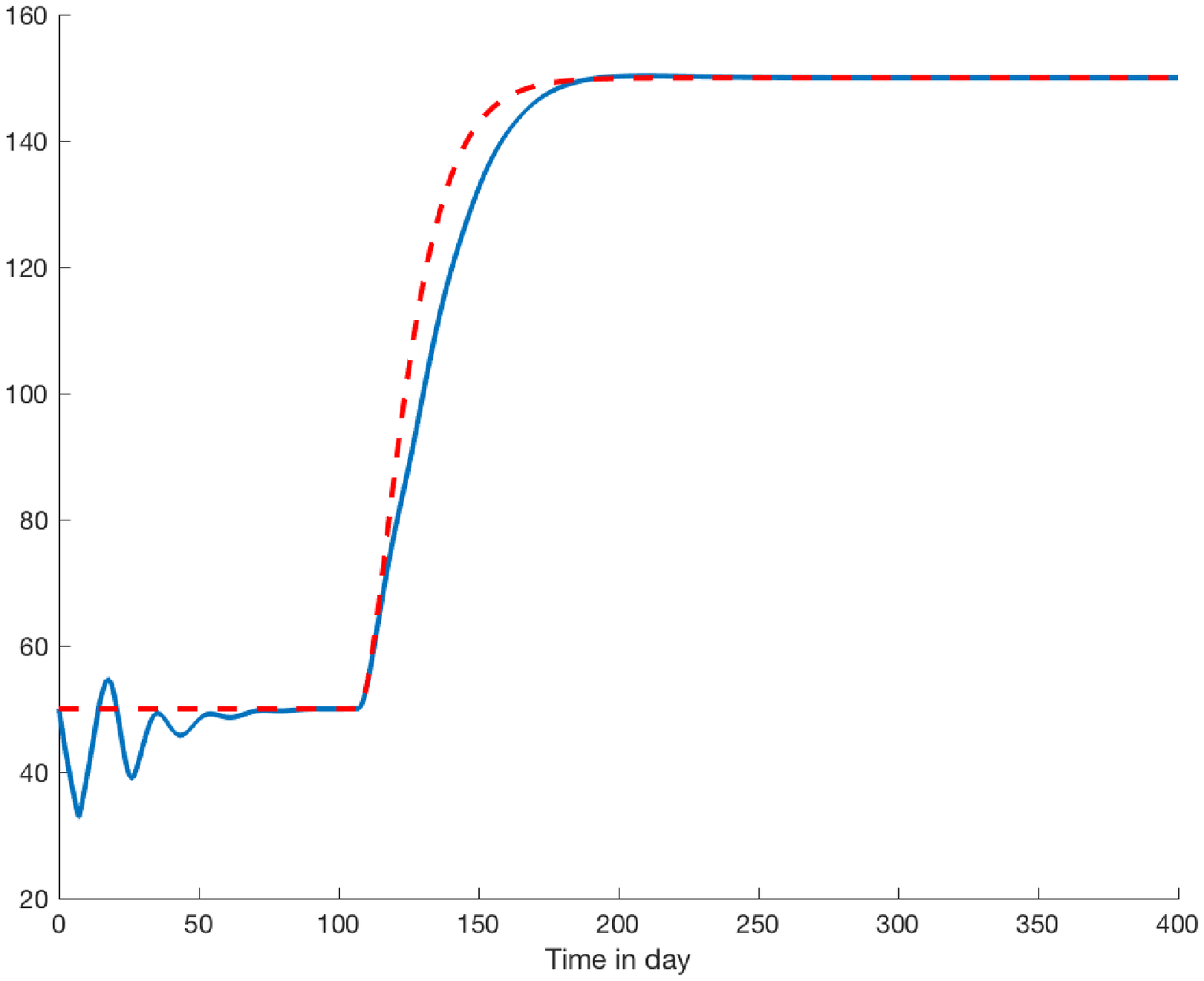,width=0.28\textwidth}}
%
%
\caption{Scenario 5}\label{S5}
\end{figure*}

\begin{figure*}[!ht]
\centering%
\subfigure[\footnotesize Control]
{\epsfig{figure=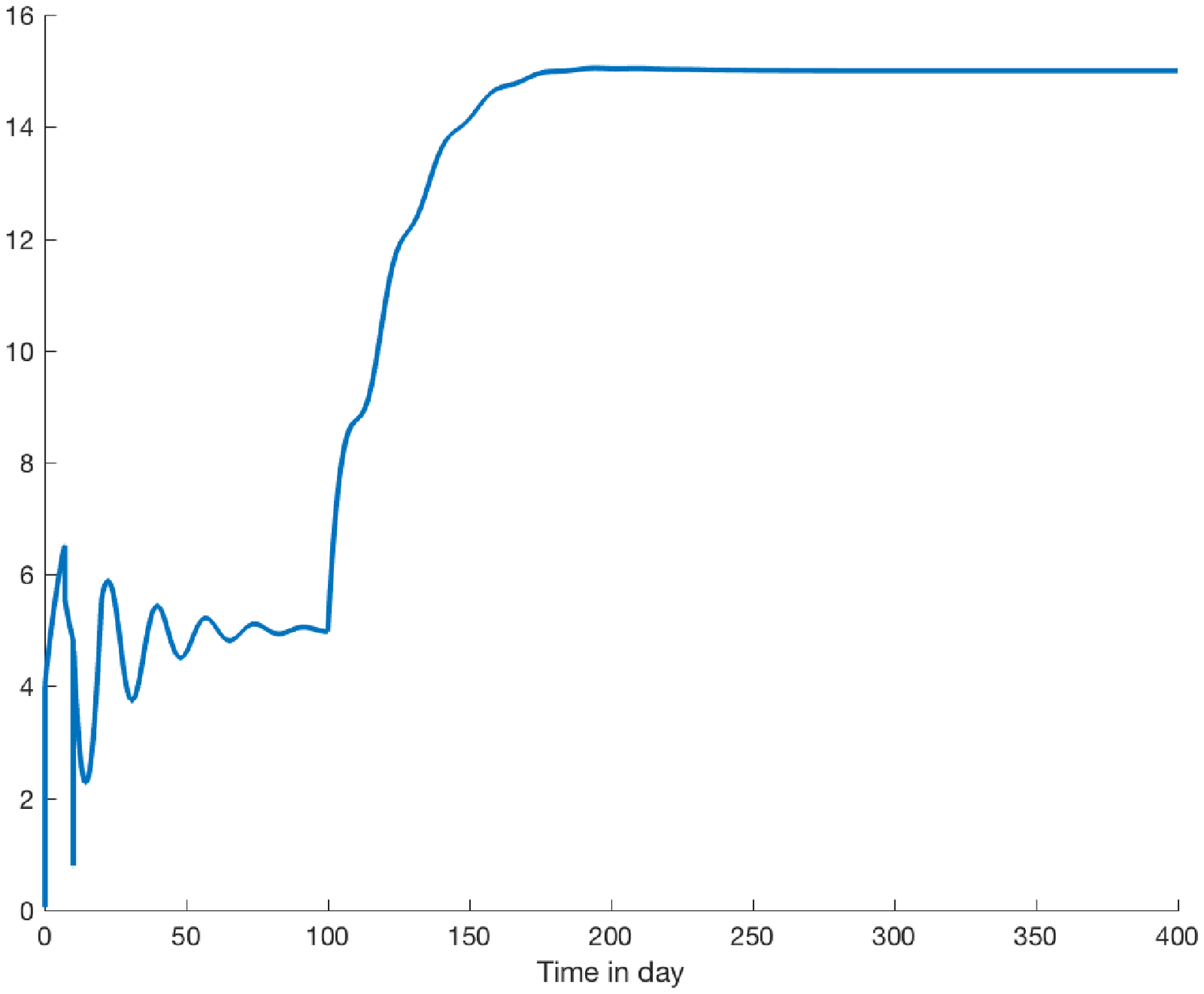,width=0.28\textwidth}}
\subfigure[\footnotesize Output (--) and Reference (- -)]
{\epsfig{figure=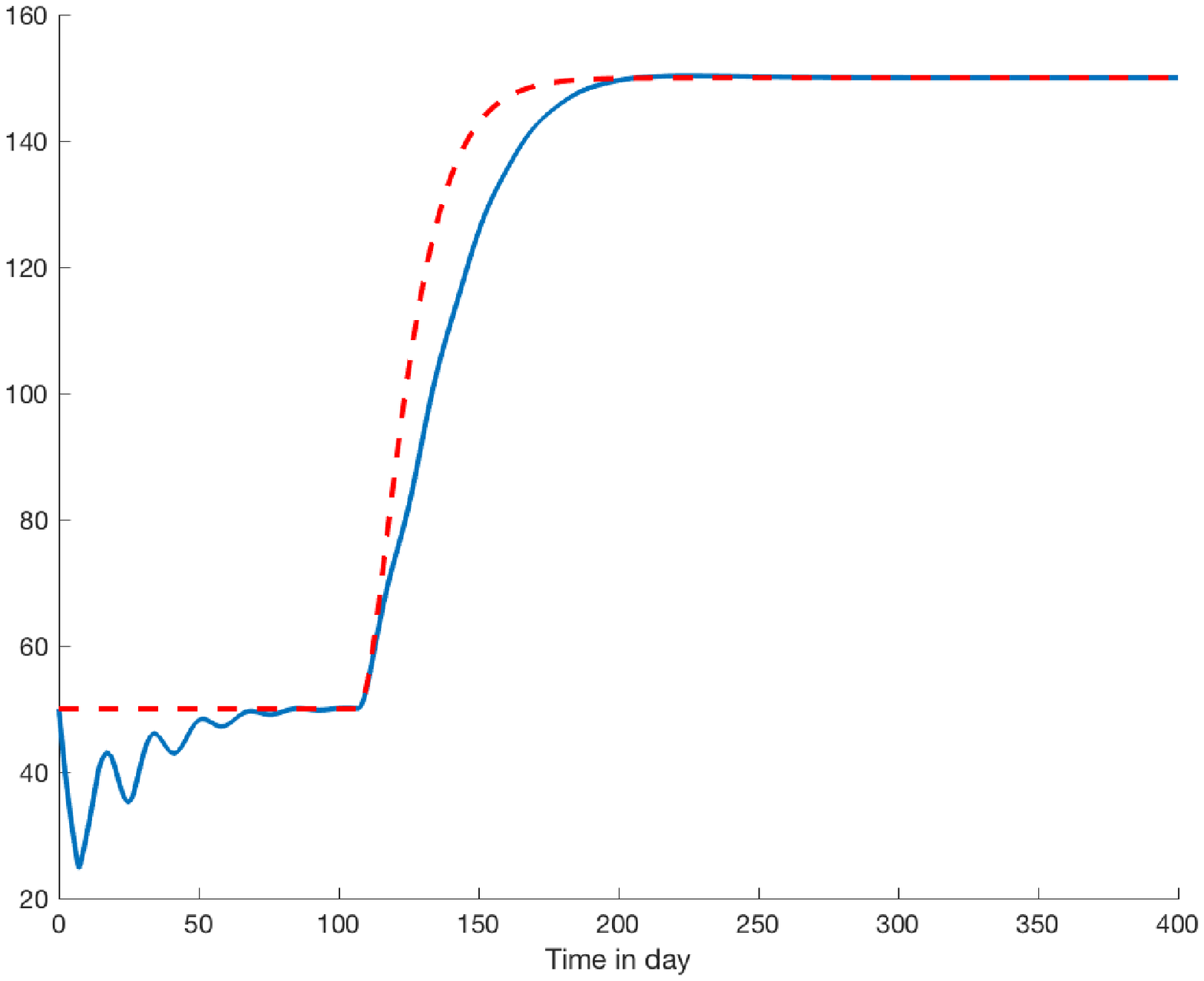,width=0.28\textwidth}}
%
%
\caption{Scenario 6}\label{S6}
\end{figure*}

 Start with $d(t)=0$ in Figures  \ref{S4}, \ref{S5}, \ref{S6}. The excellent results in Figure \ref{S4} lead to $\alpha = 1$, $K_p = 0.1$. 
 
  Inventory tracking is impressive. In order to check robustness, we selected respectively $\sigma=0.06$, $\sigma=0.1$ in Figures \ref{S5}, \ref{S6}. The outcomes remain remarkable.

\begin{figure*}[!ht]
\centering%
\subfigure[\footnotesize Control]
{\epsfig{figure=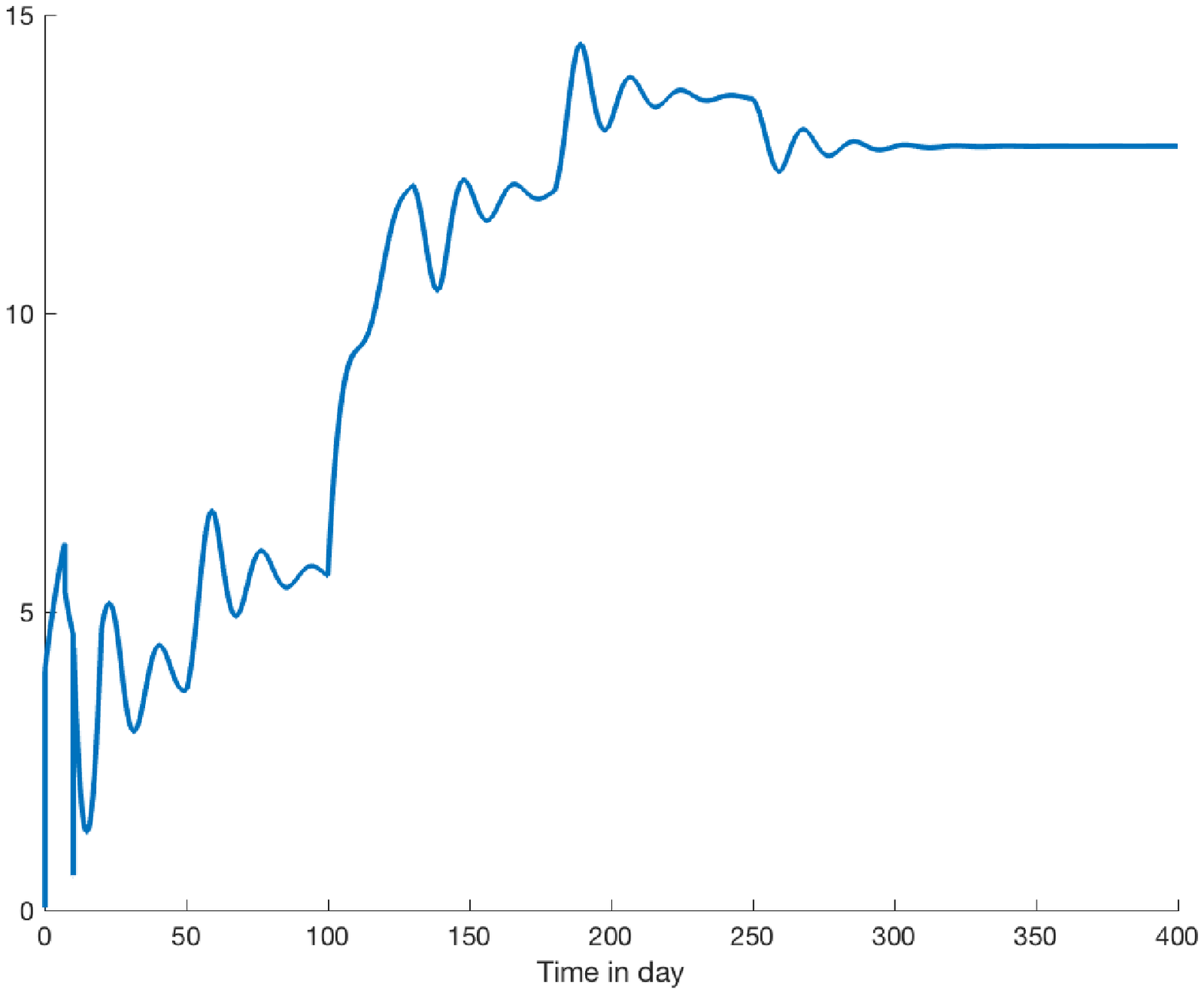,width=0.28\textwidth}}
\subfigure[\footnotesize Output (--) and Reference (- -)]
{\epsfig{figure=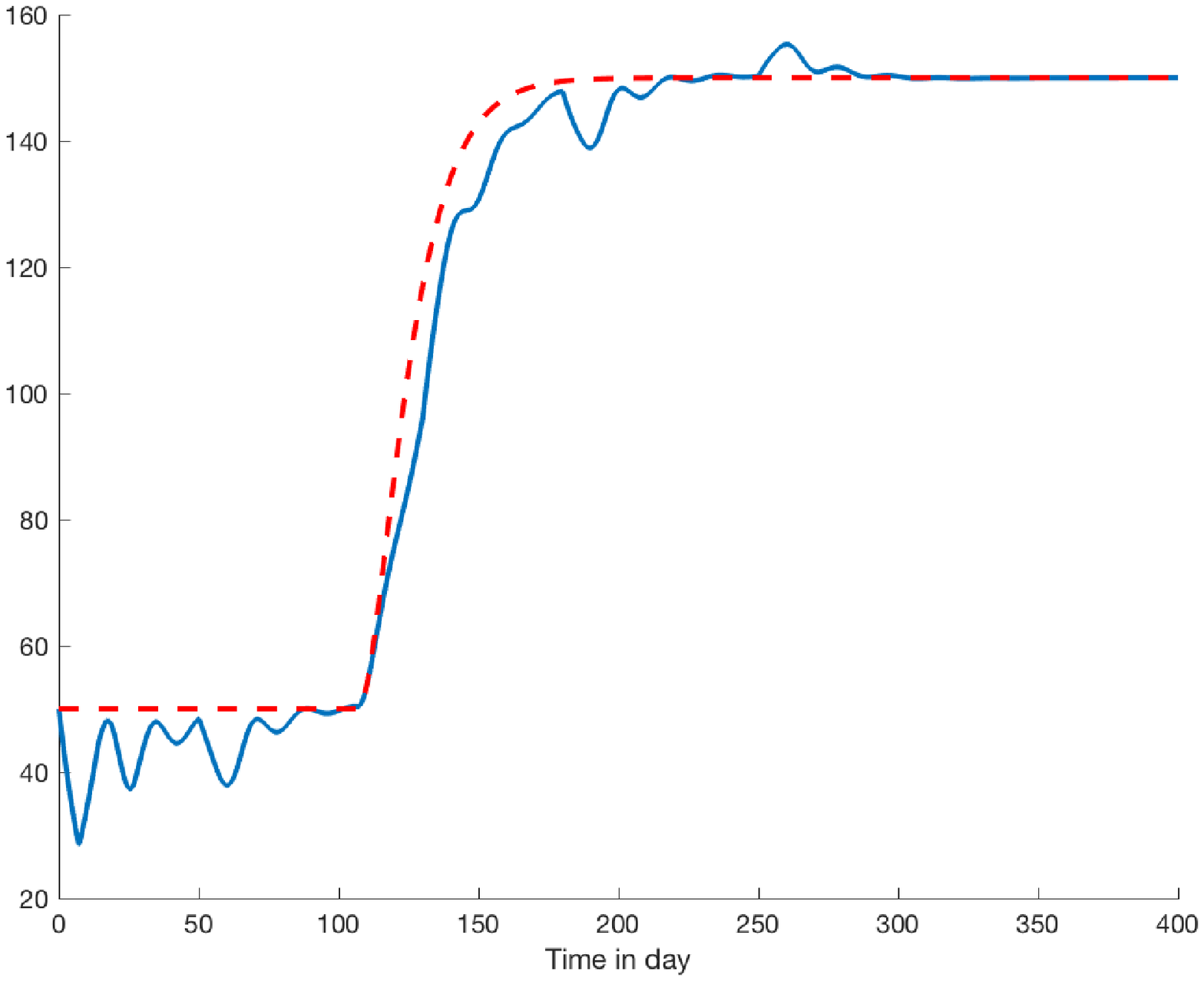,width=0.28\textwidth}}
\subfigure[\footnotesize Demand]
{\epsfig{figure=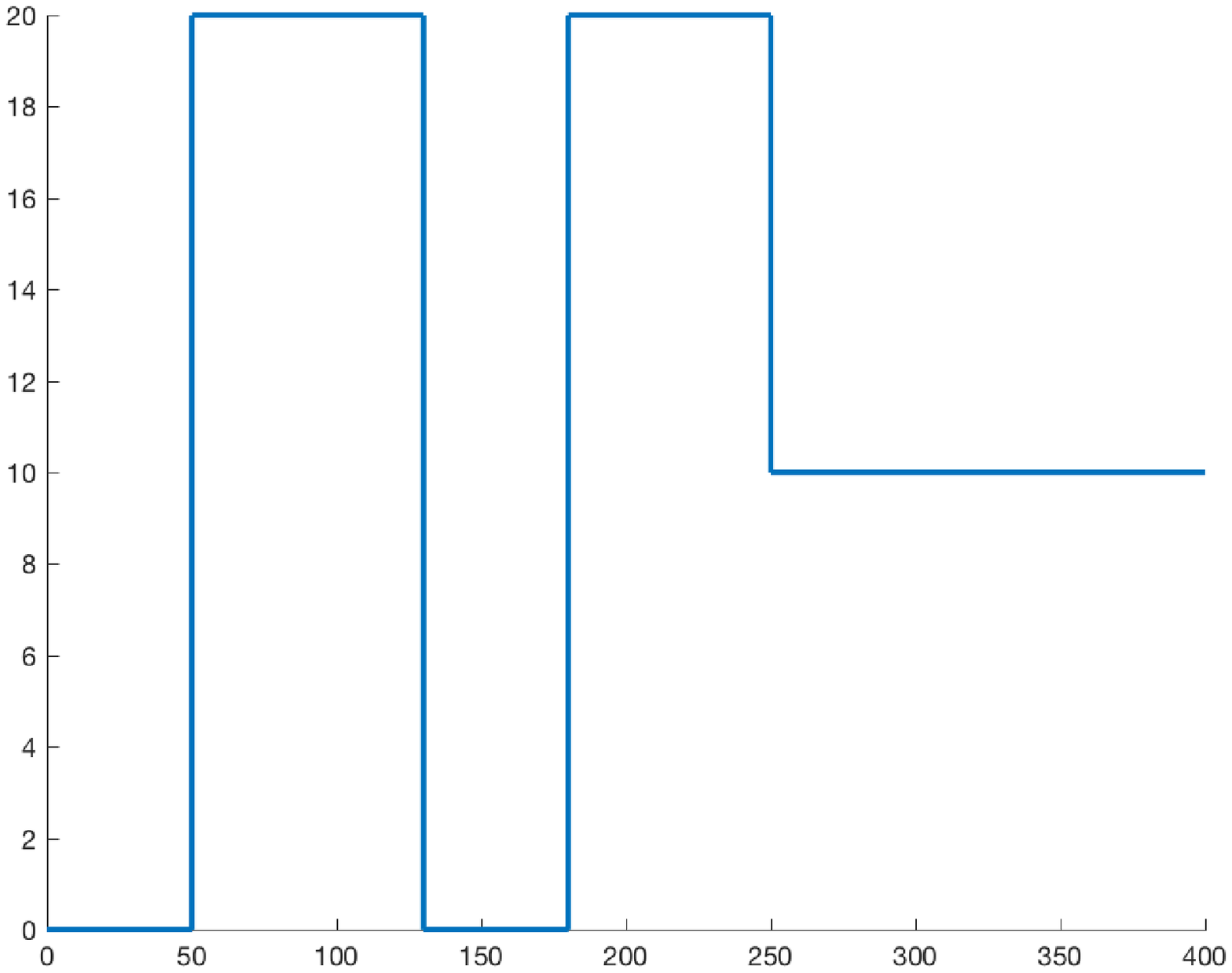,width=0.28\textwidth}}
\caption{Scenario 7}\label{S7}
\end{figure*}

In Figure \ref{S7}-(a) a demand with violent changes is introduced (see Fig. \ref{S7}-(c)). The inventory tracking is still exceptional.

\begin{figure*}[!ht]
\centering%
\subfigure[\footnotesize Control]
{\epsfig{figure=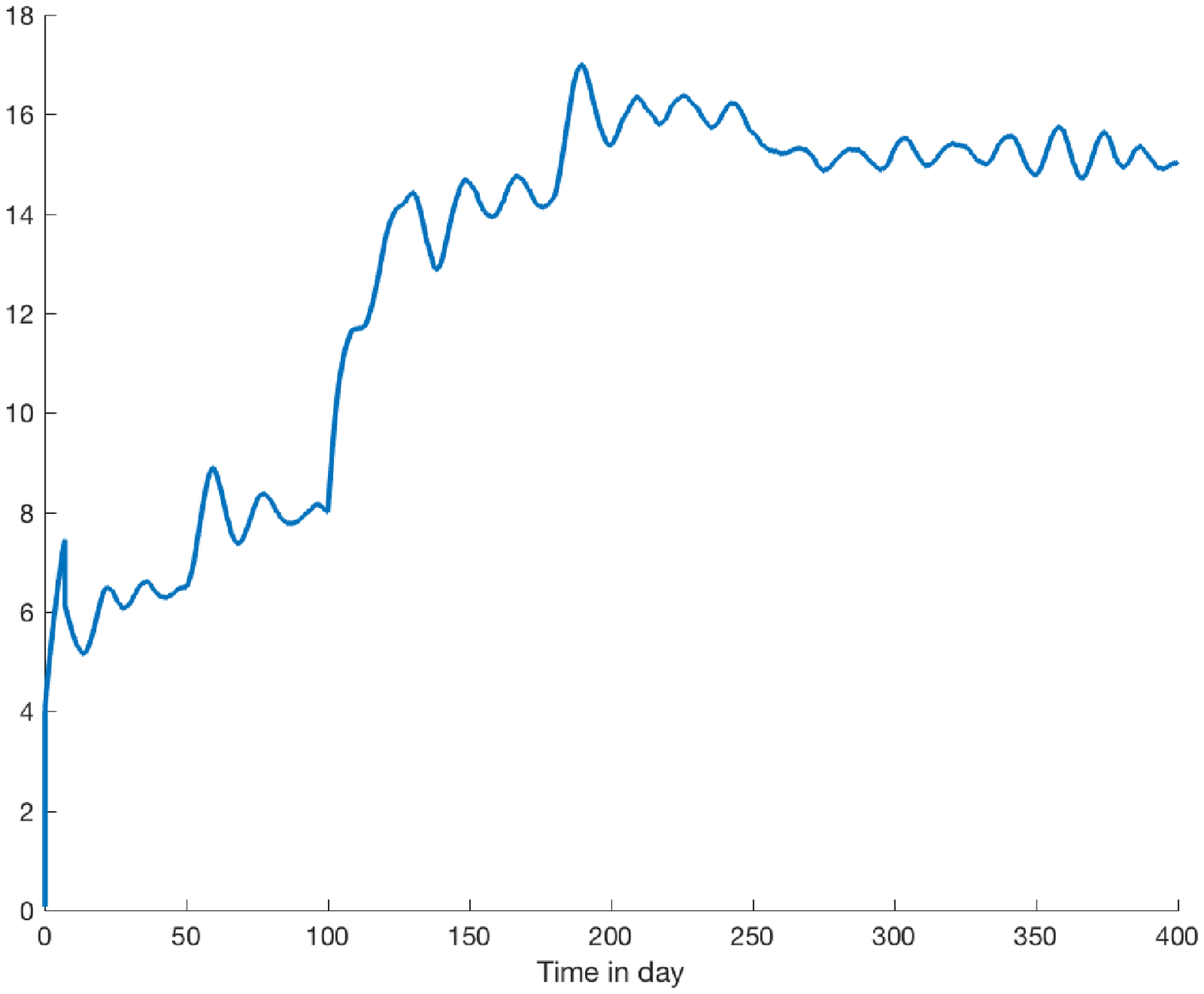,width=0.28\textwidth}}
\subfigure[\footnotesize Output (--) and Reference (- -)]
{\epsfig{figure=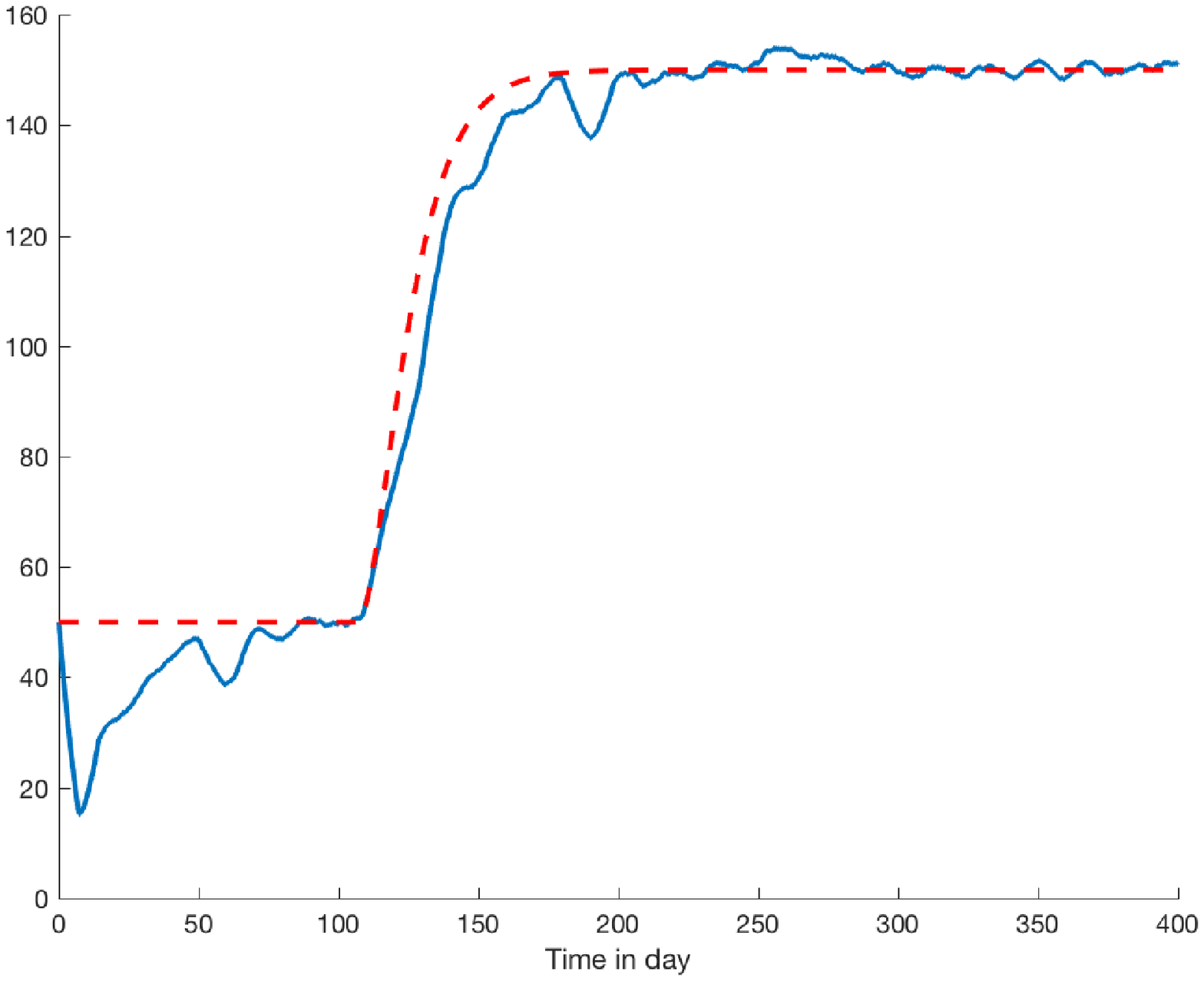,width=0.28\textwidth}}
\subfigure[\footnotesize Demand]
{\epsfig{figure=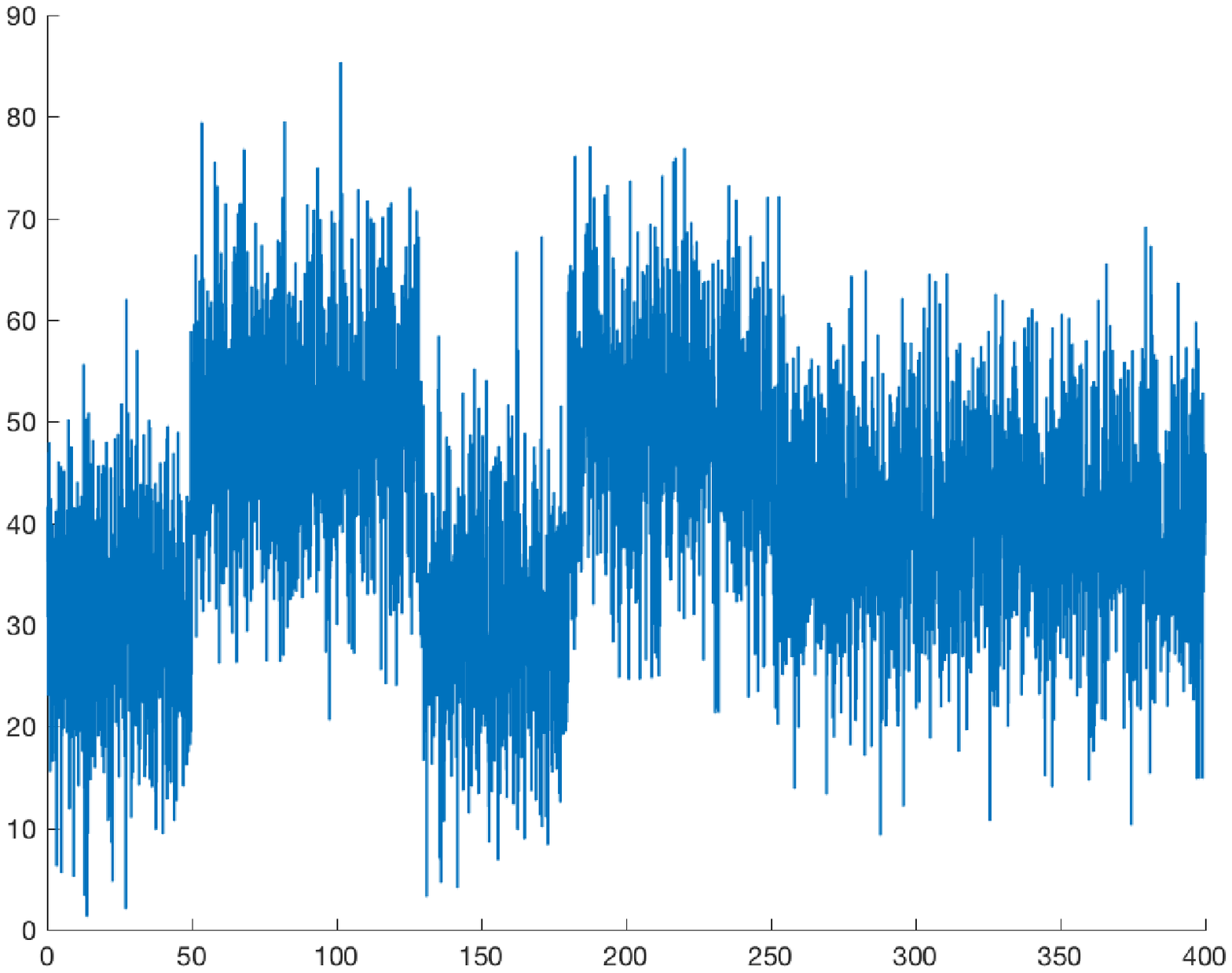,width=0.28\textwidth}}
\caption{Scenario 8}\label{S8}
\end{figure*}

In Figure \ref{S8} a corrupting Gaussian white noise, with a standard deviation equal to $10$, is taken into account (see Fig. \ref{S8}-(c)). In order to avoid negative values for $u(t)$, $d(t)$ oscillates around the value $30$. The results are still convincing.

\begin{remark}
Similar results to those in Section \ref{bull} on the bullwhip effect could be extended here without any difficulty. 
\end{remark}

\section{Conclusion}\label{conc}
A true application to concrete situations is necessary of course in order to confirm the above approach. As in financial engineering \emph{risk analysis} would also be a stimulating topic in inventory control \cite{risk}: the tools developed in \cite{sol,agadir} might lead to a solution. Related means \cite{sol} should also lead to appropriate metrics (see, \textit{e.g.}, \cite{trapero} and the references therein).

From a control-theoretic viewpoint, a completely new approach combining Smith predictors, short-term forecasting and model-free control has been presented in order to provide an adequate regulation of some delay systems. Let us remind \cite{ijc13} the reader about the severe difficulties that are due to delays in the model-free setting. Compare with other techniques in the literature: \cite{nankin,edf,doublet1,doublet2,han,huba,mustafa,thabet,yaseen,safe,neuro}.



%

\end{document}